\documentclass[a4paper,fleqn,usenatbib,useAMS]{mnras}

\usepackage{color}
\usepackage{graphicx}
\usepackage{amsmath}	% Advanced maths commands
\usepackage{amssymb}	% Extra maths symbols
\usepackage{multicol}        % Multi-column entries in tables
\usepackage{bm}		% Bold maths symbols, including upright Greek
\usepackage{pdflscape}	% Landscape pages
\usepackage{subcaption}
\usepackage{comment}
\usepackage[T1]{fontenc}
\usepackage{ae,aecompl}
\usepackage{enumitem}
\usepackage{colortbl}

\usepackage{newtxtext,newtxmath}
\usepackage{etoolbox}

\title{Subhalo sinking and off-center massive black holes in dwarf galaxies} 

\author[P. Boldrini, R. Mohayaee, J. Silk]{Pierre Boldrini$^{1}$\thanks{Contact e-mail: \href{mailto:boldrini@iap.fr}{boldrini@iap.fr}}, {Roya Mohayaee$^{1}$}, {Joseph Silk$^{1,2,3}$}
\\
$^{1}$Sorbonne Universit\'e, CNRS, UMR 7095, Institut d'Astrophysique de Paris, 98 bis bd Arago, 75014 Paris, France\\
$^{2}$Department of Physics and Astronomy, The Johns Hopkins University, Baltimore MD 21218, USA\\
$^{3}$Beecroft Institute for Particle Astrophysics and Cosmology, Department of Physics, University of Oxford, Oxford OX1 3RH, UK}
\date{In original form 2019 December 1.}

\pubyear{2019}

\begin{document}
\label{firstpage}
\pagerange{\pageref{firstpage}--\pageref{lastpage}}
\maketitle

\begin{abstract}
Using fully GPU $N$-body simulations, we demonstrate for the first time that subhalos sink and transfer energy via dynamical friction into the  centres of dwarf galaxies. This dynamical heating kicks any central massive black hole (MBH) out to tens of parsecs, especially at early epochs ($z$=1.5-3). This mechanism helps explain the observed off-center BHs in dwarf galaxies and also predicts that off-center BHs are more common in higher mass dwarf galaxies since dynamical friction becomes significantly weaker and  BHs take more time to sink back  towards the centres of their host galaxies. One consequence of off-center BHs during early epochs of dwarf galaxies is to quench  any BH feedback.  
\end{abstract}


%%%%%%%%%%%%%%%%%%%%%%%%%%%%%%%%%%%%%%%%%%%%%%%%%%

%%%%%%%%%%%%%%%%% BODY OF PAPER %%%%%%%%%%%%%%%%%%

% The MNRAS class isn't designed to include a table of contents, but for this document one is useful.
% I therefore have to do some kludging to make it work without masses of blank space.

\section{Introduction}

Most galaxies are  known to harbour  supermassive black holes (SMBHs), formed within a billion years after the Big Bang. %They reside in the centres of present day galaxies with masses of $\sim 10^{6}-10^{10}$ M$_{\sun}$ based on observations of high-redshift quasars (see \cite{2013ARA&A..51..511K} for a review). The radio source Sgr $A^{*}$ the Milky Way Galaxy is an evidence for the presence of a SMBH at its centre \citep{2019A&A...629A..32I}. 
Moreover, dwarf galaxies may frequently host massive black holes (MBHs) at their centres according to X-ray observations, among others (see \cite{2019arXiv191109678G} for a recent review). These MBHs are in the mass range $\sim 10^{3}-10^{5}$ M$_{\sun}$ and are expected to play key roles in SMBH formation scenarios that invoke galaxy mergers. 

Intriguingly, some observations of active galactic nuclei (AGN) in dwarf galaxies claim that MBHs are not located at the centers of their host galaxies. This offset varies between tens of parsecs to a few kiloparsecs \citep{2014ApJ...796L..13M,2016ApJ...817..150M,2019arXiv190904670R,2019ApJ...885L...4S}. Different scenarios have been proposed to explain these off-center BHs. Plausibly, the offset could be due to the presence of a binary system before the merger 
%or the binary BH coalescence 
(e.g.  \cite{2010PhRvD..81j4009S} and references therein), or via tidal stripping during mergers (see \cite{2018ApJ...857L..22T} and references therein), the incomplete MBH inspiraling phase of the two merging galaxies \citep{2009ApJ...690.1031B,2014ApJ...789..112C}, or the recoil of merging BHs \citep{2005LRR.....8....8M,2005MNRAS.358..913V,2007PhRvL..99d1103L,2012AdAst2012E..14K}. Furthermore, recent simulations show that BHs in dwarf galaxies are expected to be wandering around the central regions after the occurrence of mergers or due to tidal stripping \citep{2019MNRAS.482.2913B,2019MNRAS.486..101P,2010ApJ...721L.148B}. The merging mechanism seems important for blue dwarfs, whereas the old dwarfs dominate \citep{2019arXiv191108497K}. Moreover, major mergers of dwarf galaxies seems also very rare after $z\sim3$ \citep{2018MNRAS.479..319F}. As the frequencies of dwarf galaxy mergers and MBH binaries are uncertain, we propose below an alternative explanation for off-center MBH.

The cold dark matter (CDM) paradigm predicts that a very large number of dark matter substructures exist inside galactic halos \citep{2008Natur.454..735D,2008MNRAS.391.1685S}. Recently, Gaia DR2 data has provided additional evidence for these substructures \citep{2019arXiv191102662B}.
%including SMBHs, which are assumed to be the result of mergers of smaller black holes. 
Dark matter (DM) halos are growing with time, either through mergers with DM halos or by accretion of smaller halos. The latter, considered as DM subhalos, have crossed the virial radius of a larger halo at some point in the past. Subhalos interact gravitationally with all the components of the galaxy before becoming remnants of disrupted halos \citep{2019arXiv190711775Z}. In the central regions, the MBH dominates the central mass content of the galaxy \citep{2000ApJ...539L...9F,2000ApJ...539L..13G}. This is the reason why passages of subhalos near  the central regions of the host galaxy can lead to energy exchange with MBHs in dwarf galaxies. 

In this Letter, we show that subhalo crossings during their infall phase can heat the central regions of dwarf galaxies and kick the central MBH on average out to tens of parsecs from the galaxy centre over a significant fraction of  the dwarf history. Assuming average initial conditions for the subhalos, we performed $N$-body simulations with GPUs, which allow parsec resolution, to study this heating process that naturally creates off-center MBHs in dwarf galaxies. The paper is organized as follows. Section 2 provides a description of the $N$-body modelling of the dwarf galaxy and its subhalos, along with  details of our numerical simulations. In Section 3, we present our simulation results, and Section 4 discusses the implications of off-center BHs for the cusp-core problem. Section 5 presents our conclusions.

\section{Dwarf galaxy-subhalo simulation}

\begin{figure}
\centering
\includegraphics[width=0.47\textwidth]{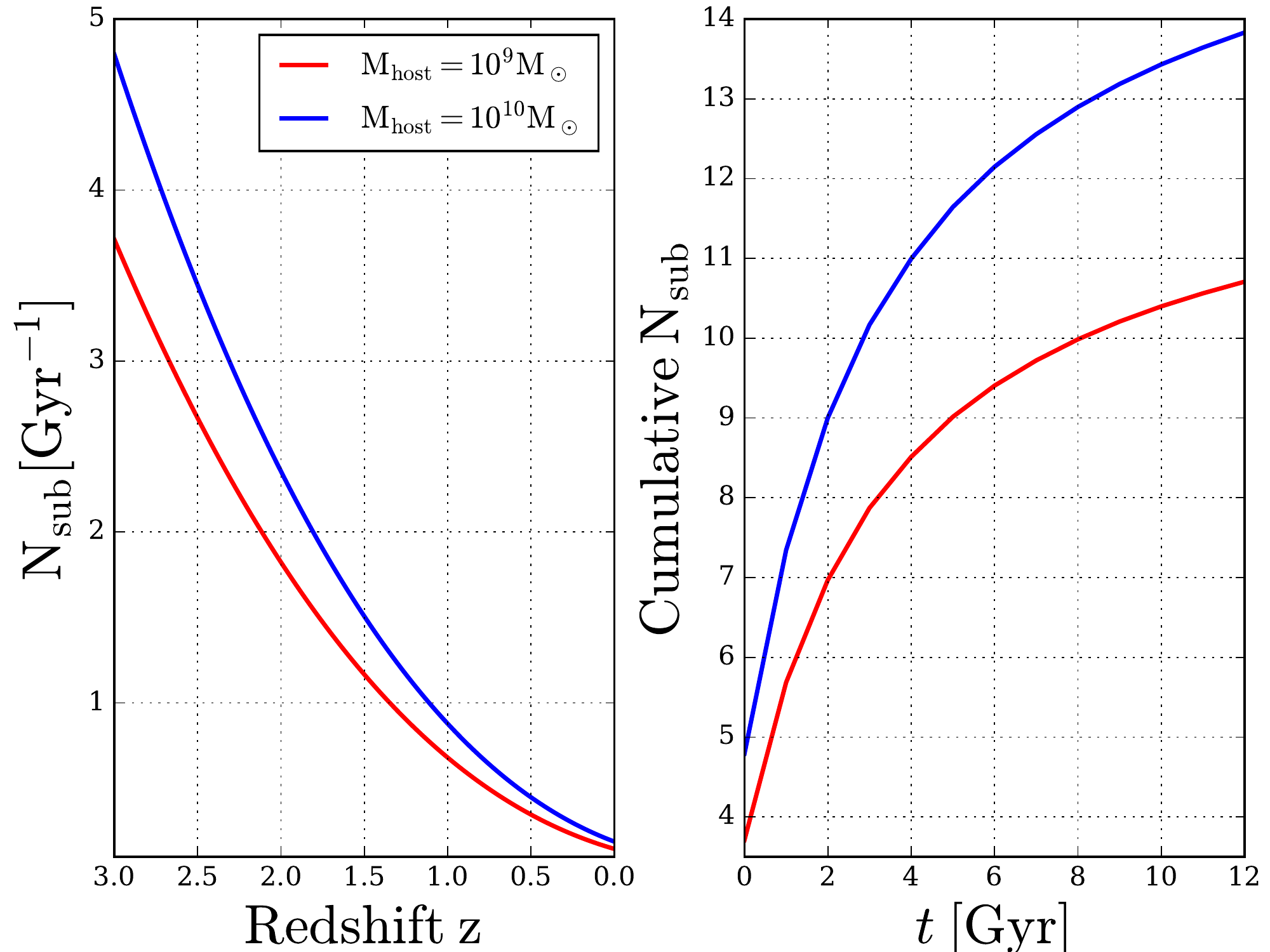}
\caption{{\it Subhalo accretions:} Average number of subhalos per Gyr as function of redshift ({\it left panel}) and cumulative average number of subhalos ({\it right panel}) as function of time with a mass ratio $10<M_{\mathrm{host}}/M_{\mathrm{sub}}<100$ in $10^9$ and $10^{10}$ M$_{\sun}$ DM host halos. This estimate is an average of the number of mergers based on the extended Press-Schechter (EPS) formalism \citep{2008MNRAS.388.1792N}.}
\label{fgR0}
\end{figure}

As our host galaxy, we consider a dwarf galaxy  that has accreted DM subhalos with  mass ratio $10<M_{\mathrm{host}}/M_{\mathrm{sub}}<100$. We construct a live dwarf galaxy composed of a central MBH ($10^{5}-10^{6}$ M$_{\sun}$), with only stars and DM particles, since dwarf galaxies contain little or no gas today. The stellar component is modelled by a Plummer profile \citep{1911MNRAS..71..460P}:
\begin{equation}
\rho(r)=\frac{3a^{2}M_{0}}{4\pi}(r^{2}+a^{2})^{-5/2},
\label{plu}
\end{equation}
where $a$ and $M_{0}$ are the scale parameter and the mass, respectively. We assume an average half-light radius of 294 pc and a mass of $10^{7}$ $M_{\sun}$ for the stellar component of the dwarf galaxy based on  Table 1 of \cite{2019MNRAS.484.1401R}. 
For the host halo and subhalos, we assume a NFW density profile \citep{1996ApJ...462..563N}: 
\begin{equation}
\rho_{\mathrm{NFW}}(r) = \rho_{0}\left(\frac{r}{r_{\mathrm{s}}}\right)^{-1}\left(1+\frac{r}{r_{\mathrm{s}}}\right)^{-2},
\label{eqn1}
\end{equation}

with  central density $\rho_{0}$ and scale-length $r_{\mathrm{s}}$. For the simulations, we consider DM host halos of $M_{\mathrm{host}}=10^9$ and $10^{10}$ M$_{\sun}$ at redshift $z=3$. Given the halo mass and redshift, both halo concentrations $c_{200}$ can be estimated from cosmological $N$-body simulations \citep{2014MNRAS.441.3359D}. The abundance of subhalo accretion for a specific host halo mass range can be determined by the extended Press-Schechter (EPS) formalism \citep{1991ApJ...379..440B,1993MNRAS.262..627L}. Fig.~\ref{fgR0} represents the average number of subhalos per Gyr as function of redshift ({\it left panel}) and cumulative average number of subhalos ({\it right panel}) as function of time with  mass ratio $10<M_{\mathrm{host}}/M_{\mathrm{sub}}<100$ in $10^9$ and $10^{10}$ M$_{\sun}$ DM host halos. We determine this rate from analytic merger rates for DM halos within the EPS formalism \citep{2008MNRAS.388.1792N}. The left panel in Fig.~\ref{fgR0} shows that a $10^9$ ($10^{10}$) M$_{\sun}$ host halo has accreted on average 3-4 (4-5) subhalos per Gyr for the adopted mass ratio at $z=3$ (see Fig.~\ref{fgR0}). Moreover, galaxies  continuously accrete smaller halos. Over their history, $10^9$ ($10^{10}$) M$_{\sun}$ host halos have accreted 10-11 (13-14) subhalos with a mass ratio $10<M_{\mathrm{host}}/M_{\mathrm{sub}}<100$ (see {\it right panel} in Fig.~\ref{fgR0}). In the simulations, the subhalo position was drawn randomly under the requirement that the initial separation between the centre of the galaxy and subhalos is the virial radius of the host halo, $r_{\mathrm{vir}}$. The subhalo orbit has an initial circularity $\eta$ depending on the host halo mass and redshift. As the orbital distributions of subhalo circularity are given to good approximation by \cite{2011MNRAS.412...49W}, we determine the average circularity $\eta=0.52$ ($\eta=0.47$) at $z=3$ for a $10^9$ ($10^{10}$) M$_{\sun}$ host halo. Here, the MBH is represented by an additional particle of mass $10^{-4}$ and $10^{-3}$ M$_{\mathrm{host}}$, placed initially at the center of the dwarf galaxy. We assume also in this study that our dwarf galaxy is in isolation.

To generate our live objects, we use the initial condition code \textsc{magi}. Adoption of a distribution-function-based method ensures that the final realization of the galaxy is in dynamical equilibrium \citep{2018MNRAS.475.2269M}. We perform our simulations with the high performance collisionless $N$-body code \textsc{gothic}, which runs entirely on GPUs \citep{2017NewA...52...65M}. We evolve the dwarf galaxy-subhalo system over 12 Gyr for $10^9$ and $10^{10}$ M$_{\sun}$ host halos. We performed simulations for 1 and 4 subhalos with mass ratios $M_{\mathrm{host}}/M_{\mathrm{sub}}=12.5$ and 50, as limiting cases. We set the particle resolution of all the live objects to 100 M$_{\sun}$ and the gravitational softening length to 2 pc. We also assess the impact of numerical effects on BH dynamics by running simulations for three different softening lengths, $\epsilon$ = 4, 2 and 1 pc. We apply these tests to the simulation over 12 Gyr for a $10^9$ M$_{\sun}$ DM host halo hosting a $10^5$ M$_{\sun}$ BH and accreting one DM subhalos with a mass of $8\times10^7$ M$_{\sun}$. Our simulations are well converged for $\epsilon=1$ and 2 pc. Our system was centered on the mass center of the stellar component for all our results. The system corresponds to all particles (DM, stars and MBH) in the simulation. The reason why we centered on the stellar component is precisely because observations establish displacements of MBH from the stellar component.

\section{Results}

\begin{figure}
\centering
\includegraphics[width=0.47\textwidth]{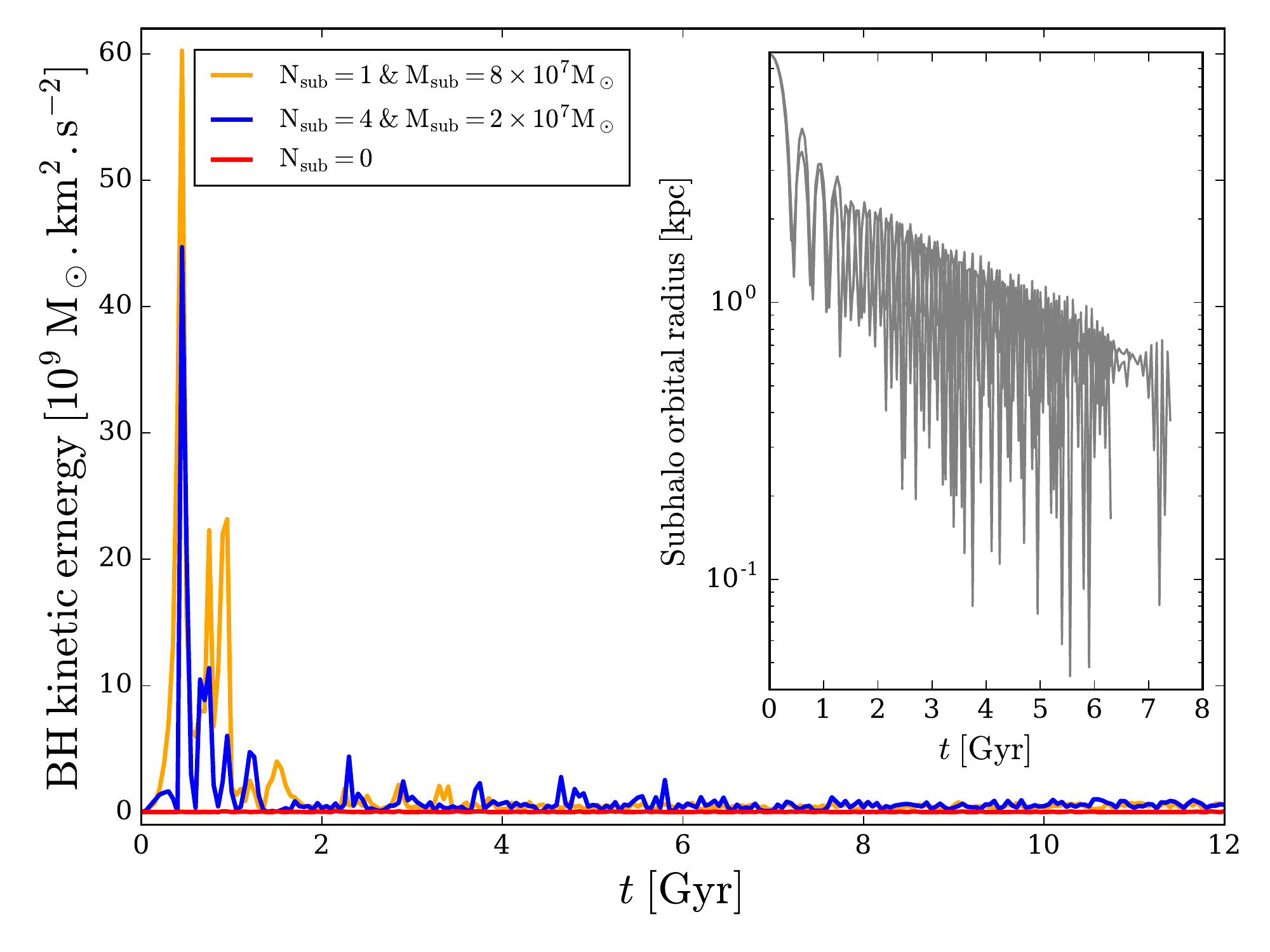}
\caption{{\it Energy transfer via dynamical friction:} Kinetic energy over 12 Gyr gained by the MBH in different scenarios (see Table~\ref{tab1} for details). As subhalos reach their velocity peak during its first infall in the galaxy, the energy transfer is maximal at their first pericentre of subhalos. Inset figure: Orbital decay of the four subhalos with a mass of $8\times10^7$ M$_{\sun}$ accreted by a dwarf galaxy embedded in a $10^9$ M$_{\sun}$ host halo over 12 Gyr. These radii correspond to the distance between the subhalo and the centre of the dwarf stellar component. The initial separation between the centres of the galaxy and subhalos is the virial radius of the host halo, $r_{\mathrm{vir}}$. Dynamical friction induced by the DM field is responsible for the infall of these subhalos. Thus, the central region of the galaxy experienced multiple subhalo crossings.}
\label{fgR1}
\end{figure}

\begin{figure}
\centering
\includegraphics[width=0.47\textwidth]{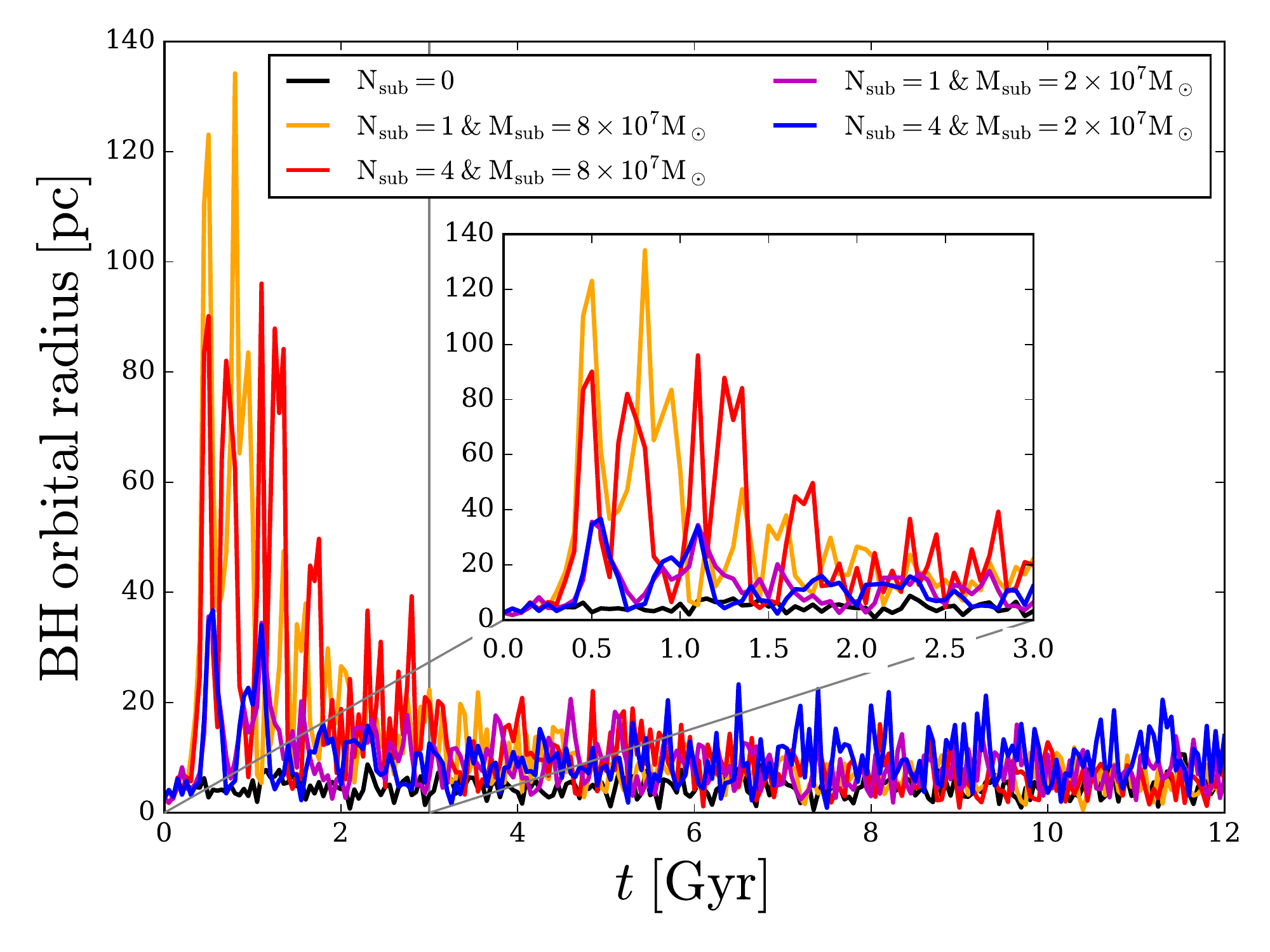}
\caption{{\it Off-center MBH:} BH orbital radius over 12 Gyr for different subhalo numbers and masses (see Table~\ref{tab1} for details). This radius corresponds to the distance between the BH and the mass centre of the dwarf stellar component. The reason why we centered on the stellar component is precisely because observations establish displacements of MBH from the stellar component. The MBH of mass $10^5$ M$_{\sun}$ is initially at the centre of the dwarf galaxy. Subhalo crossings  heat the central regions and more particularly affect the MBH via dynamical friction. Indeed, subhalos add energy to the MBH, causing it to leave the galaxy centre. In the absence of perturbers, the MBH remains at the centre of the dwarf galaxy. This scenario ensures the stability of the BH against numerical effects (black curve). However, taking into account the subhalo interactions results in a kick of the MBH to tens of parsecs from the galaxy centre, depending sensitively on the number of subhalos and on their masses. We assume that the MBH is off-center when its orbital radius is greater than the orbital radius of the MBH in the absence of  any perturbers ($N_{\mathrm{sub}}=0$)}
\label{fgR2}
\end{figure}

We consider the accretion of DM subhalos by a dwarf galaxy, which hosts a central MBH. Details of all our scenarios are given in Table~\ref{tab1}. The inset plot in Fig.~\ref{fgR1} depicts the orbital decays of four subhalos with a mass of $8\times10^7$ M$_{\sun}$ by a a dwarf galaxy embedded in a $10^9$ M$_{\sun}$ DM halo over 12 Gyr. These radii correspond to the distance between the subhalo and the centre of the dwarf stellar component. Dynamical friction induced by the DM field is responsible for the infall of these subhalos. As a result, the central region of the galaxy experiences multiple subhalo crossings (see in Fig.~\ref{fgR1}). Indeed, DM subhalos are extended objects following a NFW profile with a scale radius of $\sim$ 600 pc and their outerparts interact with the host galaxy's centre during crossings. Furthermore, subhalos also experience tidal disruptions from the dwarf galaxy. As shown in the inset plot in Fig.~\ref{fgR1}, all subhalos are completely disrupted after 6-8 Gyr. 

Subhalo crossings heat the central region and more particularly the MBH via dynamical friction. Indeed, subhalos add energy to the BH, causing it to leave the galaxy centre. Fig.~\ref{fgR2} illustrates the orbital radius of a $10^{5}$ M$_{\sun}$ MBH, initially at the galaxy centre, over 12 Gyr. This distance corresponds to the distance between the BH and the mass centre of the dwarf stellar component. In the absence of perturbers such as subhalos, the MBH remains at the centre of the dwarf galaxy. This scenario ensures the stability of the BH against numerical effects (black curve in Fig.~\ref{fgR2}). However, taking into account the subhalo interactions results in a kick of the MBH to tens of parsecs from the galaxy centre.
Indeed, the MBH has gained kinetic energy via dynamical friction. As subhalos reach their velocity peak during its first infall in the galaxy, the energy transfer is maximal at their first pericentre of subhalos (see Fig.~\ref{fgR1}). Fig.~\ref{fgR2} also highlights that the displacement of the MBH depends strongly on the number of subhalos and their masses. We assume that the MBH is off-center when its orbital radius is greater than the mean displacement of the MBH calculated over 12 Gyr in the absence of any perturbers ($N_{\mathrm{sub}}=0$) and this phenomenon is characterized by the offset time $T_{\mathrm{offset}}$. Table~\ref{tab1} shows that the MBH is off-center most of the time in all scenarios. We also calculate  the time spent by the MBH at $r_{\mathrm{BH}}>\mathrm{15\;pc}$. Before kicking the MBH, subhalos need a characteristic time to transfer energy to the central BH, defined to be the heating time $T_{\mathrm{heating}}$. The maximum offset reached by the MBH due to heating from subhalos is between 35 (run12) and 134 pc (run18) depending on the scenarios (see Table~\ref{tab1}). According to Fig.~\ref{fgR0}, $10^9$ M$_{\sun}$ DM halos accrete on average 3-4 DM subhalos at redshift z=3. Based on run42 and run48, we predict that the MBH will spend on average 1.9-2.5 Gyr beyond 15 pc due the crossings of the four subhalos after the heating time (0.25-0.35 Gyr). Thus, we expect that MBH are off-center for a significant time during the early epochs (z=1.5-3) of dwarf galaxies. Furthermore, as the subhalo accretion is a continuous process in galaxies, we expect  that most of the time, MBH are offset from the galaxy centre due to repeated heating by subhalo crossings. At low redshift, 1-2 DM subhalos are still be accreted by dwarf galaxies (see Fig.~\ref{fgR0}) and we expect that most of MBHs will be off-center by tens of parsecs even if these substructures are less concentrated at this recent epoch (see run12 and run18 in Table~\ref{tab1}). Hovewer, our mechanism cannot explain the large displacement of MBHs observed for nearby dwarf galaxies \citep{2019arXiv190904670R}.

\begin{figure}
\centering
\includegraphics[width=0.47\textwidth]{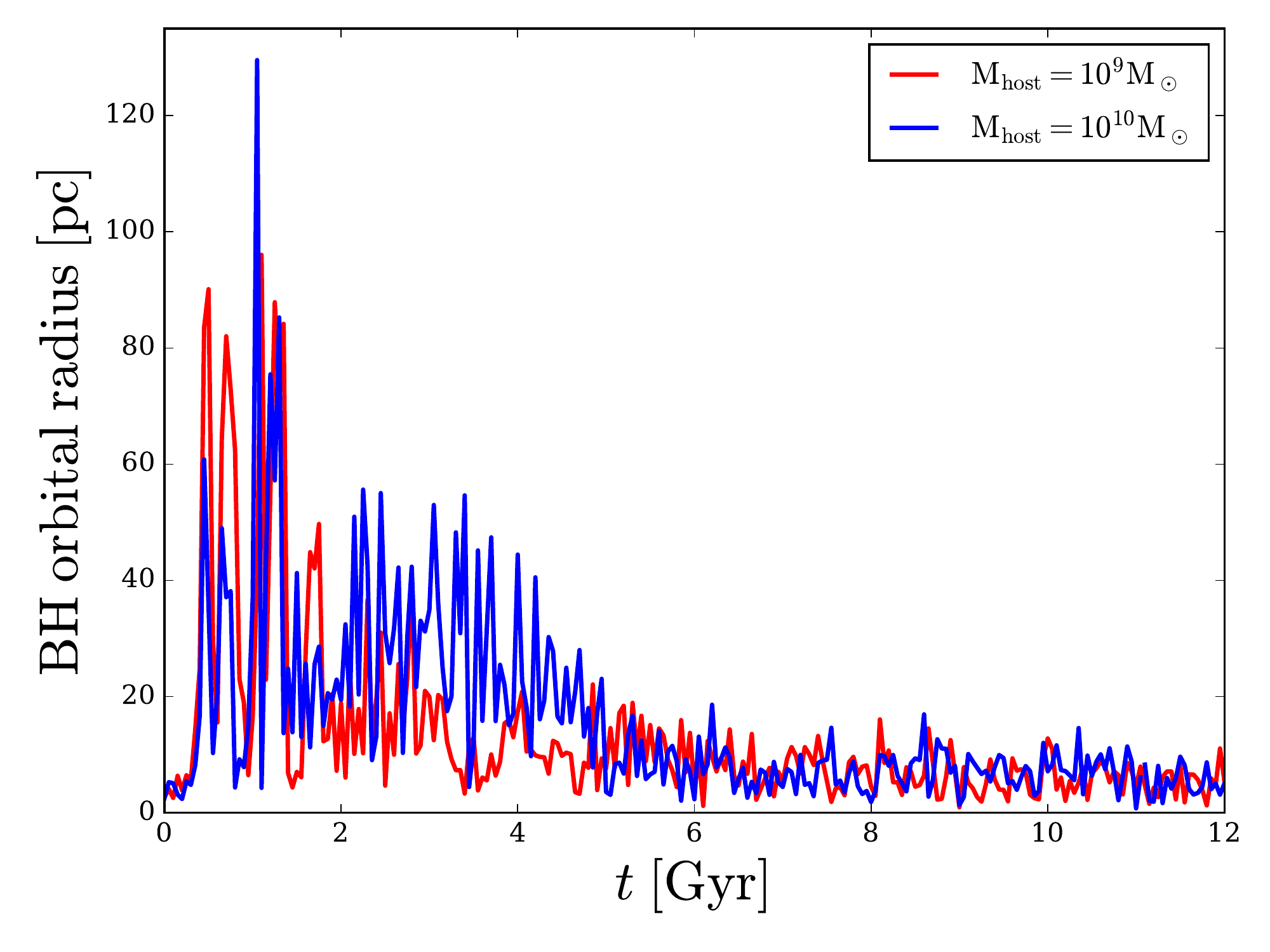}
\caption{{\it Host halo mass impact:} BH orbital radius over 12 Gyr in $10^9$ and $10^{10}$ M$_{\sun}$ DM host halos accreting four DM subhalos with a mass of $8\times10^7$ M$_{\sun}$. MBHs are off-center more time in higher mass DM host. Indeed, the MBH spent on average 3.8 Gyr beyond 15 pc after the heating phase in a $10^{10}$ M$_{\sun}$ host halo (see Table~\ref{tab1}). We predict that MBHs are going to be off-center for a longer time in higher mass galaxies.}
\label{fgR3}
\end{figure}

\begin{figure}
\centering
\includegraphics[width=0.47\textwidth]{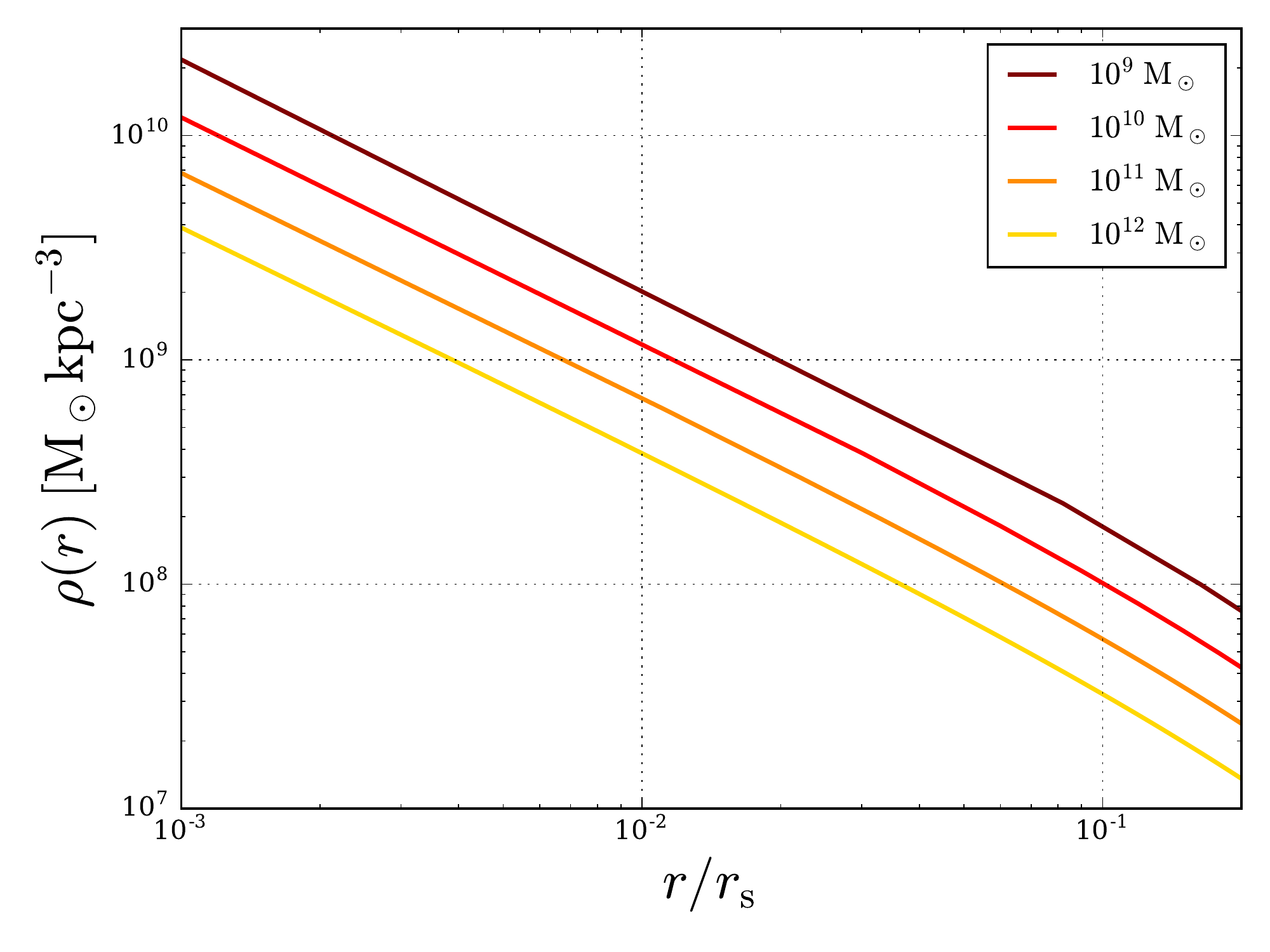}
\caption{{\it Dynamical friction in halo centres:} Density profiles of DM host halos with different masses as function of the radius normalized by their scale radius, assuming a NFW profile at redshift $z=0$. The offset time is related to the efficiency of dynamical friction, which strongly depends on the DM density at the galaxy centre. The density at the centre decreases as the halo mass grows. Consequently, the dynamical friction on BHs becomes significantly weaker and BHs take more time to sink towards the centre of higher mass galaxies after the kick.}
\label{fgR4}
\end{figure}

According to this scenario, we expect that the host halo and MBH masses play an important role. Indeed, this heating mechanism is based on the efficiency of dynamical friction, which strongly depends on the DM density at the galaxy centre. Run48 and run48b confirm that increasing the BH mass reduces its offset time because more massive objects fall in more rapidly due to dynamical friction (see Table~\ref{tab1}). We explored the host halo mass impact in Fig.~\ref{fgR3} by comparing the MBH offset induced by subhalos in  $10^9$ M$_{\sun}$ and  $10^{10}$ M$_{\sun}$ host halos over 12 Gyr. Fig.~\ref{fgR3} shows that MBH are off-center for a longer  time in higher mass DM hosts. Indeed, the MBH spends on average 3.8 Gyr beyond 15 pc after the heating phase in a $10^{10}$ M$_{\sun}$ host halo. According to our simulation results, we predict that MBHs are going to be off-center for a longer time in higher mass galaxies. This offset time is directly related to dynamical friction, which strongly depends on the DM density profile. Assuming a NFW profile, we determine the density profiles of DM host halos with different masses at redshift $z=0$ in Fig.~\ref{fgR4}. Given the halo mass and redshift, both halo concentrations $c_{200}$ can be estimated from cosmological $N$-body simulations \citep{2014MNRAS.441.3359D}.
We show that the density at the centre decreases as the halo mass grows (see Fig.~\ref{fgR4}). At high redshift, the difference between the central density as function of the DM halo mass is reduced but the same trend is respected. Consequently, we predict that off-center BHs are more common in higher mass galaxies because after the kick, dynamical friction on BHs becomes significantly weaker and then BHs take more time to sink towards the centre of these galaxies. Moreover, in high mass galaxies, MBHs are going to have less inertia due to the lower galaxy potential and thus they will reach farther distances as demonstrated in Fig.~\ref{fgR3}. Our result reinforces our prediction of a population of wandering black holes, particularly in higher mass galaxies \citep{1994MNRAS.271..317G,2002ApJ...571...30S,2003ApJ...582..559V,2004MNRAS.354..427I,2011MNRAS.414.1127M,2014ApJ...780..187R,2019MNRAS.482.2913B}.

\begin{table*}
\begin{center}
\label{tab:landscape}
\begin{tabular}{cccccccccccc}
 \hline
Simulation & $N_{\mathrm{sub}}$ & $M_{\mathrm{BH}}$ & $M_{\mathrm{sub}}$ & $M_{\mathrm{host}}$ & $M_{*}$ & m & $\epsilon$ &  $T_{\mathrm{heating}}$ &  $T_{\mathrm{offset}}$ &  $T(r_{\mathrm{BH}}>\mathrm{15\;pc})$ & $r^{\mathrm{BH}}_{\mathrm{max}}$\\
 & & [M$_{\sun}$] & [M$_{\sun}$] & [M$_{\sun}$] & [M$_{\sun}$] & [M$_{\sun}$] & [pc] & [Gyr] & [Gyr] & [Gyr] & [pc] \\
    \hline \\
    run42 & 4 & $10^{5}$ & 2$\times10^{7}$ & $10^{9}$ & $10^{7}$ & 100 & 2 & 0.35 & 10.7 & 1.9  & 37 \\
    run48 & 4 & $10^{5}$ & 8$\times10^{7}$ & $10^{9}$ & $10^{7}$ & 100 & 2 & 0.25 & 9.4 & 2.5  & 96\\
    run12 & 1 & $10^{5}$ & 2$\times10^{7}$ & $10^{9}$ & $10^{7}$ & 100 & 2 & 0.4 & 9.95 & 1.15  & 35\\
    run18 & 1 & $10^{5}$ & 8$\times10^{7}$ & $10^{9}$ & $10^{7}$ & 100 & 2 & 0.3 & 9.5 & 2.2  & 134\\ 
    run48b & 4 & $10^{6}$ & 8$\times10^{7}$ & $10^{9}$ & $10^{7}$ & 100 & 2 & 0.25 & 8.25 & 1.9  & 92\\
    run48m & 4 & $10^{5}$ & 8$\times10^{8}$ & $10^{10}$ & $10^{7}$ & 100 & 2 & 0.3 & 9.95 & 3.8 & 129\\
    
    \hline
\end{tabular}
\caption{Simulation parameters for all the scenarios. From left to right, the columns give: the number of subhalos; the MBH mass; the subhalo mass; the DM host halo mass; the stellar mass; the mass resolution; the softening length; the heating time; the offset time; the time spent by the MBH at $r_{\mathrm{BH}}>\mathrm{15\;pc}$; the maximal distance reached by the MBH. The maximum offset reached by the MBH due to heating from subhalos is between 35 and 134 pc. MBHs will spend on average 1.9-2.5 Gyr beyond 15 pc due the crossings of the four subhalos after the heating time (0.25-0.35 Gyr). Thus, we expect that MBHs are off-center during a significant time in the early epoch (z=1.5-3) of dwarf galaxies.}
\label{tab1}
\end{center}
\end{table*}

\section{The cusp-core problem}

One of the key predictions of the $\Lambda$ cold dark matter ($\mathrm{\Lambda CDM}$) model of cosmology is that DM assembles into halos that, in the absence of baryon effects, develop cuspy density profiles following the Navarro-Frenk-White (NFW) form. This important result was obtained from $N$-body simulations and is independent of initial conditions and cosmological parameters \citep{1996ApJ...462..563N,1997ApJ...490..493N}. Nevertheless, the latter study shows that the DM density profile seems not to be universal \citep{2010MNRAS.402...21N}. In contrast, the DM density profiles of dwarf galaxies, inferred from their HI rotation curves or stellar kinematics, reveal shallower profiles than the NFW model and that are consistent with a central density core \citep{1994Natur.370..629M,1994ApJ...427L...1F,1995ApJ...447L..25B,2001AJ....122.2396D,2011MNRAS.414.3617K,2013MNRAS.433.2314H,2015AJ....149..180O}. This discrepancy between theory and observation corresponds to the so-called core-cusp problem, considered to be  one of the greatest challenges faced by the CDM paradigm. In order to resolve this discrepancy, many mechanisms involving baryons have been proposed, which could transform cusps into cores via changes in the gravitational potential caused by stellar feedback redistributing gas clouds, generating bulk motions and galactic winds along with heating by dynamical friction of massive clumps \citep[e.g.][]{2011ApJ...736L...2O,2013MNRAS.429.3068T,2012MNRAS.421.3464P,2001ApJ...560..636E,2010ApJ...725.1707G}.

Solutions invoking baryonic feedback cycles can potentially reconcile observed dwarf galaxy anomalies with $\mathrm{\Lambda CDM}$ predictions. This challenge at small scales occurs precisely where baryons play an important role, notably through BH feedback that generates significant movements of the gas. BH feedback can expel large amounts of gas from the central   of galaxies. A fraction of this gas then cools and returns to the centre, generating repeated cycles of significant outflows which in turn cause rapid fluctuations of the gravitational potential. These potential fluctuations dynamically heat the DM and lead to the formation of a core \citep[e.g.][]{2013MNRAS.432.1947M,2017MNRAS.472.2153P,2017ApJ...839L..13S}. The gradually dispersion of the DM particles away from the center of the halo is ultimately responsible for core creation.

Numerical simulations show that the peak of AGN activity happens between $z\sim3$ and $z\sim1.6$. We sould be able to observe the flattening of the DM density profile induced by high BH activity. Especially during this early phase of galaxy evolution, we predict that MBHs are off-center according to our simulation results. Due to the heating by subhalos, we have shown that BHs remain on average tens of parsecs away throughout most of the halo’s history. However, BHs accrete gas inefficiently away from the galaxy centre as gas clumps are centrally located \citep{2018MNRAS.480.3762S}. Then, the conditions required to alter the deep potential of galaxies appear to be missing. Without BH feedback, the inner density profiles of DM halos will remain cuspy. In addition, off-center BHs entail the quenching of BH feedback in dwarf galaxies. Consequently, it seems difficult to induce DM core formation in dwarfs from BH feedback. Baryonic feedback cycles are the preferred option \citep{2012MNRAS.421.3464P}.

\section{Conclusions}

We have shown that the heating of the central region in dwarf galaxies by subhalos via dynamical friction entails the offset of MBHs, especially at  early epochs (z=1.5-3). Indeed, at redshift z=3, the average number of subhalo accretions is high ($\sim$4 per Gyr) and then the sinking of subhalos transfers energy to the galaxy centre and especially to the MBH, causing it to leave the central region. The heating by subhalos and the subsequent kick to the central MBH provides a new mechanism that  contributes to explain observed off-center BHs in dwarf galaxies. We have also predicted that off-center BHs are more common in higher mass galaxies because after the kick, dynamical friction on BHs becomes significantly weaker and then BHs take more time to sink towards the centre of these galaxies. As BH feedback consists of energy injection into halos, this latter is commonly invoked as a mechanism for core formation. Indeed, BH feedback can induce cusp-to-core transition for the DM halo. Here, we have argued that the main consequence of off-center BHs during early epochs of dwarf galaxies is the quenching of BH feedback and then the absence of DM core formation by this mechanism. 

Dynamical perturbations induced by subhalo crossings, causing MBHs to vacate the galaxy center, can also modify the spatial distribution of the other galaxy components such as stars and DM particles. Stars heated by subhalos can contribute to populating the stellar halo  as an alternative to star formation in gas outflows that are associated with starburst activity \citep{2017Natur.544..202M,2019MNRAS.485.3409G,2019MNRAS.486..344R,2019arXiv191203316Y}. A notable difference between these scenarios will be the age disdtribution  of the ejected stars. This mechanism will be studied in a forthcoming publication.

%\section{Acknowledgments}
%We thank Miki Yohei to provide us the non-public $N$-body code, \textsc{gothic}. We would like also to thank Apolline %Guillot, Dante von Einzbern and George Boole for their constructive suggestions on improving the manuscript. 

%%%%%%%%%%%%%%%%%%%%%%%%%%%%%%%%%%%%%%%%%%%%%%%%%%

% Don't change these lines
\bsp	% typesetting comment
\label{lastpage}
\end{document}